\documentstyle[12pt,world_sci]{article}
\newcommand{\beq}[1]{\begin{equation} \label{eq:#1}}
\newcommand{\eeq}{\end{equation}}
\newcommand{\abs}[1]{\mid \! #1 \! \mid}
\newcommand{\avg}[1]{< \! #1 \! >}

\newcommand{\eqref}[1]{(\ref{eq:#1})}
\newcommand{\phitil}{{\tilde{\phi}}}
\newcommand{\Nmin}{N_{min}}
\newcommand{\Nrun}{N_{run}}

\begin{document}

\hspace*{4.0in}UCD-92-17\\
\hspace*{4.5in}July 1992\\
\title{{\bf FINITE SIZE SCALING OF PROBABILITY DISTRIBUTIONS IN  SU(2)
LATTICE GAUGE THEORY AND $\phi^4$ FIELD THEORY}
\vspace*{0.3cm}}

\author{STUART STANIFORD-CHEN\\
\vspace*{0.3cm}
	{\em Department of Physics, University of California, Davis, \\
	California 95616, USA\\
	stuarts@jeeves.ucdavis.edu}}

\maketitle

\begin{center}
\parbox{13.0cm}
{\begin{center} ABSTRACT \end{center}
{\small \hspace*{0.3cm}

For a system near a second order phase transition, the probability distribution
for the order parameter can be given a finite size scaling form.  This fact is
used to compare the finite temperature phase transition for the Wilson lines in
$d=3+1$ SU(2) lattice gauge theory with the phase transition in $d=3$ $\phi^4$
field theory.  I exhibit the finite size scaled probability distributions in
the form of a function of two variables (the reduced `temperature' and the
magnetization) for both models.  The two surfaces look identical, and an
analysis of the errors also suggests that they are the same.  This strengthens
the idea that the SU(2) effective line theory is in the Ising universality
class.  I argue for the wider application of the method used here.}}
\end{center}


\section{Overview}
\label{sec:overview}

Some time ago, Svetitsky and Yaffe \cite{finscal:SY_orig,finscal:SY_physrep}
conjectured that the phase transition undergone by the Wilson lines in finite
temperature SU(2) lattice gauge theory should be in the same universality class
as the corresponding Ising model.  In particular the 3+1 dimensional SU(2)
theory should have the same exponents and so forth as the three-dimensional
Ising model.  The basis for this conjecture is no more certain than noticing
that the dimensionality and the symmetry of the Wilson line theory are the same
as those of the Ising model, convincing oneself that the interactions in the
effective line model are short range, and hoping that under renormalization
group transformations the two theories tend to the same fixed point in the
appropriate space of possible actions.

To test this idea, a considerable amount of work has been done both using
various analytic approximations and numerically.  However, to date, work has
mainly concentrated on verifying that the critical exponents of the two models
are the same.  The critical exponents of the three-dimensional Ising model are
now known fairly precisely, and measurements of the corresponding quantities
for SU(2) are in reasonably good agreement
\cite{finscal:joeMC1,finscal:joeMC2,finscal:oldeng,finscal:eng90,finscal:nweng}.
However, there are many other quantities
which should be constant across a universality class.  For instance, there are
a number of amplitude ratios which can be measured.  There are also a variety
of functions which are universal up to some rescalings.  A typical example
would be the finite size scaling form\footnote{As the author rapidly tires of
writing (or reading) `finite size scaling', this phrase will often be
abbreviated to just `finite size' in what follows.} of the susceptibility.
Some attention has been paid to these issues \cite{finscal:amp_rat}, but the
agreement between the models is only partially satisfactory.  It appears that
the small lattices to which we are currently limited in simulating the gauge
theory forbid a very convincing comparison of amplitude ratios.  The purpose of
this paper is to address any consequent uncertainty as to the universality
class of the gauge theory.

Some time ago it was noted by Bruce\cite{finscal:orig_bruce} and by
Binder\cite{finscal:orig_bind} that, in the vicinity of a second order phase
transition, the whole probability distribution for the order parameter can be
given a finite size form. The usual finite size forms of the susceptibility,
magnetization, etc. then appear as various moments of this function.  In this
paper I shall consider the finite size form of the probability distribution at
varying temperatures.  Thus the object forms a surface; the probability is a
function of the scaled magnetization and of the scaled temperature.  This
surface should be universal, and hence it is possible to compare the phase
transitions in two different models by comparing their probability surfaces.
It should be noted that the form of the surface does depend on the boundary
conditions, though not the details of the action.  In the past,
\cite{finscal:orig_bruce,finscal:orig_bind,finscal:bruce_bdr1,finscal:bruce_bdr2}
this has only been done to the
limited extent of comparing the probability distributions right at the critical
point.  Here I expand on the technique and consider the full surface in a
finite region around the phase transition.

I will show the form of this surface for SU(2) based on existing Monte Carlo
data.  I have taken Monte Carlo data for the $\phi^4$ scalar model, which
should also be in the same universality class, and I will exhibit the surface
for that model, and argue that the two are the same within the appropriate
errors.  Visually they are strikingly similar.  This provides further evidence
that SU(2) lattice gauge theory really is in the same universality class as the
Ising model and $\phi^4$ field theory.  For readers who prefer to look at
the pictures before they read the equations, the comparisons are in Figures
\ref{fig:view1} through \ref{fig:contours}.

As to the organisation of the paper: Section~\ref{sec:defns} introduces
notation and definitions; Section~\ref{sec:claim} contains the facts about the
finite size scaling surface that are needed in this paper;
Section~\ref{sec:theory} provides renormalization group arguments for the
universality of the scaling surface; Section~\ref{sec:MonteCarlo} briefly
discusses the generation of the data for $\phi^4$; Section~\ref{sec:comparison}
consists of the comparison of the surface for the two models, and
Section~\ref{sec:conclusion} concludes.


\section{Definitions}
\label{sec:defns}

In the case of $\phi^4$ field theory, we take an $N^3$ lattice with fields
$\phi_i$ living at each site.  We posit the action to be
\beq{phi4action}
S_\phi = \frac{\alpha}{2} \sum_{<ij>} (\phi_i-\phi_j)^2
       - \frac{r}{2!} \sum_{i} {\phi_i}^2
       + \frac{u}{4!} \sum_{i} {\phi_i}^4          .
\eeq
Note the sign of the second term.  We will be mainly interested in the absolute
magnetization $\phi$, which is defined by
\beq{abs_mag}
\phi = \frac{1}{N^3}\abs{\sum_{i} \phi_i}          .
\eeq
This quantity is defined at every instant, or perhaps at every sweep in a Monte
Carlo simulation.  Its thermal average is denoted by
${\cal M}_\phi = \ < \! \! \phi \! \! >$.  We can also define a corresponding
susceptibility ${\cal X}_\phi$ by
\beq{abs_susc}
{\cal X}_\phi =  N^3 ( \avg{\phi^2}- \avg{\phi}^2 ).
\eeq
In the case of the SU(2) gauge theory, we have an $N^3 \times N_\tau$ lattice
with gauge variables $U_l$ living on each link.  The Euclidean action is the
normal Wilson one:
\beq{wilson_action}
S_U = \frac{4}{g^2} \sum_\Box {(1 - \frac{1}{2} \mbox{Tr}U_\Box)},
\eeq
where $U_\Box$ is the usual plaquette variable
\beq{U_plaquette}
U_\Box = \prod_{l\in\Box}U_l .
\eeq
The physical temperature in this theory is given by
\beq{Physical_temp}
T = \frac{1}{a N_\tau} .
\eeq
We require that this temperature is finite (ie. $N_\tau < N$), but otherwise we
shall not be concerned with it.

The Wilson line observables $L_i$ are obtained by tracing over all the timelike
links at a fixed spatial position $i$:
\beq{wilson_line}
L_i = \frac{1}{2}\mbox{Tr}\prod_{\tau=1}^{N_\tau} U_{i,\tau} .
\eeq
These are defined in such a way as to be real variables taking values in
$[-1,1]$, and hence one can think of them as though they were spins.  We can
then define an absolute instantaneous average of these quantities:
\beq{abs_Lmag}
L = \frac{1}{N^3}\abs{\sum_{i}L_i}.
\eeq
By analogy with the $\phi^4$ case we also define the thermal expectation value
${\cal M}_L = \avg{L}$, and a susceptibility
\beq{abs_Lsusc}
{\cal X}_L = N^3 (\avg{L^2}-\avg{L}^2).
\eeq
For the remainder of this paper, we will take no interest in the $U_l$
variables but will instead focus entirely on the $L_i$ variables, treating them
as the site variables in some effective three-dimensional theory of which we do
not quite know the action.  It is known, however, that this theory undergoes a
second order phase transition at some critical value of the coupling $g_c$.
Therefore we define a reduced `temperature' by
\beq{su2_t}
t_L = \frac{4/g^2 - 4/{g_c}^2}{4/{g_c}^2}.
\eeq
This quantity is distinct from the physical temperature of Equation
\eqref{Physical_temp}.  Similarly, the $\phi^4$ theory has a critical point at
some $r_c$.  We define its reduced `temperature' as
\beq{phi4_t}
t_\phi = \frac{r-r_c}{r_c}.
\eeq
For convenience, in the rest of this paper we will refer to these variables as
reduced temperatures (without quotes).

Now, with our definitions, both models are in their broken (magnetized) phase
when $t$ is positive, and their unbroken phase when $t$ is negative.  We expect
the normal amplitude-exponent behaviour of the infinite system in the critical
region:
\beq{beta}
{\cal M} = {\cal M}_+ t^\beta
\eeq
for $t$ small and positive, while
\beq{gamma_def}
{\cal X} = {\cal X}_\pm \abs{t}^{-\gamma}
\eeq
for $t$ small and positive (${\cal X}_+$), or negative (${\cal X}_-$).
We also take the usual definition for the correlation length exponent $\nu$, so
that in the critical region
\beq{nu_def}
{\xi} = {\xi}_\pm \abs{t}^{-\nu} .
\eeq
Here ${\cal M}$, ${\cal X}$ and $t$ without subscripts are doing duty for both
models discussed in this work.  Potentially the exponents for the two models
could be different.  Research to date provides moderately convincing evidence
that they are the same.  The logic of this paper is to assume that the
exponents are the same and then investigate whether the other quantities that
will be introduced match as we expect.


\section{Nature of the Finite Size Scaling Surface}
\label{sec:claim}

If we look at the $\phi$ measured on one sweep of the lattice in a simulation,
it will not typically be equal to its thermal average ${\cal M}_\phi$, but
rather will be drawn from some probability distribution $P_\phi(N,t_\phi,\phi)$
which has ${\cal M}_\phi$ as its expectation value and ${\cal X}_\phi/N^3$ as
its variance.  This last is obvious from the definition of ${\cal X}_\phi$,
Equation \eqref{abs_susc}.  We can study this probability distribution by
taking a sufficiently large collection of measurements of $\phi$ and making a
normalised histogram of them.  As the notation for $P_\phi$ indicates, the form
of this histogram depends not only on the reduced temperature $t_\phi$, but
also on the size $N$ of the lattice.

The first thing we shall need is to define a finite size scaling version of
these histograms.  We take the ansatz
\beq{phi4_q}
P_\phi = N^{\beta/\nu}Q_\phi(t_\phi N^{1/\nu},\phi N^{\beta/\nu}) .
\eeq

An algorithm to obtain $Q_\phi$ by simulations is then to construct the
histograms $P_\phi$ for a range of different values of $N$ and $t_\phi$, scale
their ordinates by $N^{-\beta/\nu}$, scale their second abscissae by
$N^{\beta/\nu}$, rearrange them along the $t_\phi N^{1/\nu}$ axis, and then
interpolate between them in some appropriate manner.  The result is a surface
$Q(z_\phi,\tilde{\phi})$ (or at any rate an approximation to it).  Here we have
introduced
\beq{z_phi}
z_\phi = t_\phi N^{1/\nu} ,
\eeq
and
\beq{phi_tilde}
\tilde{\phi} = \phi N^{\beta/\nu}
\eeq
for convenience.

Of course, we can follow exactly the same procedure for the SU(2) case,
constructing the probability distributions $P_L(N,t_L,L)$ and then taking
\beq{su2_q}
P_L = N^{\beta/\nu}Q_L(z_L,\tilde{L}) ,
\eeq
where
\beq{z_L}
z_L = t_L N^{1/\nu}
\eeq
and
\beq{L_tilde}
\tilde{L} = L N^{\beta/\nu} .
\eeq
We shall refer to these surfaces $Q$ as finite size scaling surfaces.

The next thing to note is that the surface $Q(z,\tilde{\phi})$ is a universal
quantity up to rescalings of its two arguments.  In other words, all models in
a particular universality class should have a $Q$ surface of the same shape.
The next section of this paper will be devoted to justifying from a
renormalization group approach that these claims are valid.

It is straightforward to check that the truth of \eqref{phi4_q}, the finite
size form for the probability distributions, implies the usual finite size
relation for the magnetization ${\cal M}$
\beq{finite_M}
{\cal M} = N^{-\beta/\nu}Q_M(z)
\eeq
and a similar equation for the susceptibility ${\cal X}$.  It also follows that
the universality of $Q$ guarantees the universality of the finite size forms of
these quantities.


\section{Theoretical arguments}
\label{sec:theory}

Here we will justify, on the basis of real space renormalization group
arguments (block spins and so on), the statements made in the previous section.
 Really, the reasons why the probability distributions have a finite size form
are exactly the same as the reasons why anything else has a finite size form.
However, for completeness, we shall go through it. The standard textbook
treatments of the real space renormalization group are gestured at in a few
sentences and then we turn to finite size scaling per se in somewhat more
detail.

We have our lattice with fields $\phi_i$ on a lattice of size $N^d$.  We
suppose that we have divided the lattice up into some blocks of size $b$ on a
side.  The block fields are averages of the $\phi_i$ in the block, except that,
if we wish our block spin transformation to have a non-trivial fixed point, we
have to rescale them by some non-trivial factor as well.  The correct
definition to do this is\cite{finscal:parisi}
\beq{block_spin}
\psi_k = \frac{1}{b^{d - \beta/\nu}}\sum_{i \in k} \phi_i .
\eeq
Now we imagine integrating out the original $\phi_i$ variables to find the
effective theory for these block fields $\psi_k$.  The result is some different
action from the one that we had before.  We also have to add a constant to the
action to keep the energy scale from running away but this will not concern us.
 As usual, we suppose that this renormalization group transformation is
represented by some operator on the space of all possible actions for theories
with the same symmetry and dimensionality.  We further suppose that the
transformation has some fixed point, that the transformation equations are
analytic at that fixed point, and that everything that matters happens either
in the region in which the transformation equations can safely be linearized
(when the eigenvectors of the linearized transformation are the scaling
fields), or at least in a region outside that, in which the scaling fields have
become non-linear but still respond to RG transformations in the same way.  We
assume that in the linear region the reduced temperature is one of the scaling
fields, and its associated exponent is $1/\nu$.  Thus in the region we shall
assume we are in, the effect of a RG transformation on the reduced temperature
is to change
\beq{t_transf}
t \rightarrow b^{1/\nu}t .
\eeq
Strictly speaking, after enough RG transformations, we will be driven far
enough along the direction of this scaling field that it will no longer be safe
to approximate it just by $t$, but it will be some more complex function of $t$
and other parameters.  We will just call it the $t$-field, whatever it might
be.

We are interested in probability distributions for the average value of the
fields.  Now the average over the $\psi_k$ is exactly the average over the
$\phi_i$ except for the factor of $b^{-\beta/\nu}$ that appears in the
denominator in Equation \eqref{block_spin}.  We can describe this by saying
that under a renormalization group transformation
\beq{phi_transf}
\phi \rightarrow b^{\beta/\nu}\phi .
\eeq
We also note at this point that in equation \eqref{block_spin} we have arranged
matters so that the scale of the $\phi_i$ is fixed under RG transformations.
Now, for our two models to end up at the same fixed point action after many RG
transformations we must ensure that the scale of the fields in both models is
the same.  Thus we must rescale one or the other of them.  However, since the
two sets of fields enter into actions which are of superficially quite
different forms, we cannot calculate this non-universal scale factor and must
leave it arbitrary in our analysis.

The main assumption of RG derivations of finite size
scaling\cite{finscal:barber} is that the RG transformation mostly preserves the
locality of the action.  That is, if the original action only involves a few
local couplings, then only a few couplings at finite distances will be
important in the renormalized action also.  This means, and this is the crucial
point, that the form of the RG equations (for one set of couplings in terms of
the other) will be the same in the finite system as in the infinite system.
The only way this can go wrong is if the finite system is so small that it
cannot hold all the couplings needed.  This places an upper limit on the number
of times we can apply a RG transformation to a theory on a finite lattice.

So, we now have all the assumptions at hand and can begin the argument proper.
Imagine our two theories at different points in the space of all possible
actions, $S_1$ and $S_2$.  We suppose that we apply the maximum feasible number
of RG transformations to them, which will be $n$ say, so that the resulting
lattices are the same minimal size, $\Nmin$, and have actions $S_1'$ and
$S_2'$.  We assume that the original lattices were very large, and the original
theories were at their critical points.  Hence $n$ is large, and $S_1'$ and
$S_2'$ are very close to the fixed point action $S_f$.  Since these three
theories are very similar (ie.\ all their parameters are very similar), and
since they are living on a small finite lattice (of the minimal size to fit the
necessary couplings on), they all have an extremely similar probability
distribution for the average value of their fields.  Since the average value of
these fields is exactly the average value of the original fields, except for a
known scaling factor, this distribution will give us the shape function we need
for the surface at t=0.  Suppose this distribution is $P_f(\psi_n)$.  In this
instance \eqref{phi_transf} becomes
\beq{psi_scale}
\psi_n = b^{n \beta/\nu}\phi ,
\eeq
so
\beq{P_N_scale_pre}
P(N,0,\phi) = b^{n \beta/\nu}P_f(b^{n \beta/\nu}\phi).
\eeq
The $b^{n \beta/\nu}$ factor in front comes from keeping the measure straight
(or making sure $P$ and $P_f$ are both normalised).
We can rewrite this as
\beq{P_N_scale}
P(N,0,\phi) = (N/\Nmin)^{\beta/\nu}P_f((N/\Nmin)^{\beta/\nu}\phi).
\eeq
This is clearly of the form of the finite size surface \eqref{phi4_q}
(restricted to $t=0$).  The same form applies to both models equally (except
for the arbitrary scale factor mentioned earlier).  It is clear, however, that
since the origin of this shape is a theory on a small lattice, the details of
the form are likely to be dependent on the boundary conditions (as has proved
to be the case\cite{finscal:orig_bind}).

Now we must deal with $t\neq 0$ (but still in the critical region).  If we
start off with our $S_1$ and $S_2$ and start applying RG transformations then
they will initially follow trajectories roughly toward the fixed point.
However, $t$ is being increased with every transformation (ie.\ it is
relevant), so after our $n$ transformations we are very close to the line which
has all scaling fields zero except for the $t$-field.  This gets us to some
theory on that line for the minimal size lattice.  That gives us our shape
function.  We then just pull it back to our original theories.  The shape
function will be analytic in $t$ since it comes from a theory on a finite
lattice (of size $\Nmin$).  The algebra is similar to that for $\phi$ above,
and so we get
\beq{P_N_scale_full}
P(N,t,\phi) =
(N/\Nmin)^{\beta/\nu}Q((N/\Nmin)^{\beta/\nu}\phi,(N/\Nmin)^{1/\nu}t) ,
\eeq
which is the same as \eqref{phi4_q} after we absorb $\Nmin$ into the definition
of $Q$.  Where did the arbitrary rescaling of $t$ between the models come from?
 Well we made no effort to define our reduced temperature variables,
\eqref{su2_t} and \eqref{phi4_t}, in a comparable way (not that we could easily
have done so), so they will not be the same thing in terms of the $t$ in some
systematic coordinatisation of the space of actions.

We could go on and do the same for the magnetic field $h$, and we would get a
finite size scaling hyper-surface.  However, quite besides the affront to the
English language, the author was unable to locate any good four-dimensional
plotting software.  Hence all relevant variables besides $t$ are taken as zero
in this paper.


\section{Monte Carlo data}
\label{sec:MonteCarlo}

To make use of the above ideas I needed to take data for the probability
distributions for another model in the same universality class as SU(2), data
for which was already available \cite{finscal:eng90}.  For other purposes, I
had developed a multigrid Monte Carlo code for $\phi^4$ field theory in three
dimensions, so I made my comparison with that model.  The correctness of the
code was checked by tests in the Gaussian limit and by comparison with other
published results for the susceptibility \cite{finscal:weston}.

All the data reported in this paper were taken at $u=50$ and $\alpha=1$.  Data
were taken on $8^3$, $16^3$, and $32^3$ lattices with periodic boundary
conditions.  A summary of the data set is given in Table \ref{tab:phi4_data}.
The critical point was determined by a finite size comparison of the peaks of
the susceptibility to be at $r=4.2485$.  The exponents were taken to be the
Ising values.  Specifically, based on the discussion in Liu and
Fisher,\cite{finscal:liu_fisher} I took $\nu = 0.63$, and $\beta = 0.33$.  All
the histograms were then rescaled as suggested in Section~\ref{sec:claim}, and
plotted on a single graph.  The results are shown in Figure
\ref{fig:phi4_hist}.  However, it is hard to visualise data plotted in such a
manner, and still harder to compare two different sets of data.  For that
reason, I fitted an appropriate smooth function to the data.  For numerical
convenience in the fitting process it is easier to switch to using a log of the
finite size form of the histograms $Q(z,\phitil)$ introduced earlier, and so I
looked at
\beq{S_Q}
S_Q(z,\phitil) = -\ln Q(z,\phitil) .
\eeq
The form used to fit this was then
\beq{g}
g(z,\phitil) = -\frac{a(z) \phitil^2}{2!} + \frac{b(z) \phitil^4}{4!} +
\frac{c(z)\phitil^6}{6!} - d(z),
\eeq
where
\beq{a}
a(z) = a_0 + a_1 z + a_2 z^2 + a_3 z^3,
\eeq
\beq{b}
b(z) = b_0 + b_1 z + b_2 z^2,
\eeq
\beq{c}
c(z) = c_0 + c_1 z + c_2 z^2 ,
\eeq
\beq{d}
d(z) = d_0 + d_1 z + d_2 z^2 + d_3 z^3 + d_4 z^4 .
\eeq

Here $a_0$ through $d_4$ are parameters determined by minimizing the $\chi^2$
of the fit.  The form \eqref{g} through \eqref{d} is somewhat inspired by
perturbative considerations\cite{finscal:brezin}; however, in this paper I make
no attempt to connect with $\epsilon$-expansion work.  It is also the first few
terms of the most general power series possible given the global symmetry
(which dictates the use of only even powers of $\phi$).  My criterion for
selecting it is simply that it provides a fairly good fit to the data.

The process of fitting is complicated by the lack of any simple means to do an
error analysis.  There are two main problems.  The first is that the
fluctuations in the histograms from the values they would have given an
infinite data set are correlated across neighbouring bins (ie. different values
of $\phitil$).  The second is that there are systematic errors between the
histograms from different lattice sizes used in the surface, probably due to
corrections to finite size scaling (presumably both from irrelevant operators
and from analytic effects).

To make some effort to control this, I did the fits using error bars for the
data generated in two different ways.  In the first case I just gave every bin
in every histogram an error bar of equal size, arbitrarily taken to be one, and
then minimised the resulting $\chi^2$ to obtain a fit.  In the second case I
obtained error bars by fitting to each individual histogram (prior to taking
the logarithm) a form derived from the fit surface \eqref{g} by taking its
exponential and fixing $z$ (the value of which has no effect in this instance,
being absorbed into the $a0$...$d4$ variables).  These fits all had very
reasonable $\chi^2$ per degree of freedom.  Then I derived the error bars for
each bin by assuming its value was a Poisson variable with a mean given by the
value of the fit at that $\phitil$.  (This was actually repeated until fit and
error bars were self consistent). Then those error bars were used as the input
to a full fit of \eqref{g} to all the data at once (now with minus the
logarithm taken).  This second procedure has the advantage that the resultant
$\chi^2$ is a meaningful number with the usual interpretation: the $\chi^2$ per
degree of freedom is supposed to be one (assuming the errors are independent
and Gaussian).  All analyses presented here were done with both procedures and
the conclusions were not substantially different.  Although this is a somewhat
crude method, I feel it is unlikely that any feature of the treatment which
appears the same in two such different procedures will be misrepresented due to
the defects of the error analysis.  Graphs and numbers presented in this paper
are based on the second method (which did work somewhat better than the first),
and the actual parameters obtained are, for this $\phi^4$ case, given in Table
\ref{tab:phi4_coeff}.  The full data sets and error bars are available on
request from the author.

The results of this procedure are displayed in Figure \ref{fig:phi4_data}.  At
the risk of insulting the reader's intelligence I remind him or her once again
that the original histograms have by this point been rescaled by some factors
of the lattice size and have had minus their logarithm taken.  Both fit and
data are shown there, and readers with good eyesight will be able to ascertain
that the surface is generally plausible as a fit to the data.  There are some
problems, particularly for large $z_\phi$ and modest $\phitil$.  The $\chi^2$
per degree of freedom for the whole fit is $1.69$.  Clearly then there are
systematic failures of the data to conform to this surface.  Nonetheless, I
feel that these are really rather modest, particularly given that the smallest
lattices used were only $8^3$, and that Figure \ref{fig:phi4_data} does provide
fairly good evidence both for the existence of a finite size scaling surface
$S_Q$, and for the efficacy of the polynomial \eqref{g} as a fit to it.


\section{Comparison of models}
\label{sec:comparison}

For the case of the SU(2) lattice gauge theory, data is available for
$N_\tau=4$ and the work of computing the critical point and so forth has
already been done \cite{finscal:eng90}.  I simply rescaled the histograms
and fit the surface of form \eqref{g} in the same manner as described in the
previous section for the $\phi^4$ model.  The coefficients are in Table
\ref{tab:su2_coeff}.  The resulting picture is in Figure \ref{fig:su2_data}.
Again, the data appear to be making a reasonable effort to conform to a smooth
surface.  The reader has only to look at the untouched histograms in Figure 1
of the paper by Engels {\em et al\/} in which the SU(2) data originally
appeared\cite{finscal:eng90} to realise that, by contrast, they certainly do
not form a single surface in the unscaled variables ($t_L$ and $L$ rather than
$z_L$ and $\tilde{L}$).

The value of $\chi^2$ per degree of freedom for this fit of all the SU(2) data
is 3.54.  Clearly there are more significant systematic deviations from the
data in this case than in the $\phi^4$ case.  I do not have a complete
understanding of why this should be.  It may be simply that the corrections to
scaling are larger for SU(2).  An alternative explanation is that the SU(2)
data has better statistics, allowing systematic histogram to histogram
variations to stand out of the random errors more clearly.  This motivates the
following slightly speculative digression on practical matters.  The run
lengths are given as $N_{run}$ in Tables \ref{tab:phi4_data} and
\ref{tab:su2_data}.  The SU(2) numbers are indeed much larger.  On the face of
things, though, they are not directly comparable as the algorithms used to
obtain the data were not the same.  The SU(2) data were taken with a local heat
bath algorithm, while the $\phi^4$ data were obtained with a multigrid
Metropolis algorithm.  However, in practice this may make little difference.
This is because measuring a histogram requires knowing many higher moments of
the distribution to get the smooth shape right, not just the magnetization.  In
a simulation, my intuition is that these higher moments decorrelate much faster
than the low moments (that is their auto-correlation time is much lower).  This
is because a small change in the order parameter measurement typically involves
a large change in high powers of the order parameter.  Alternatively, one can
think of the high moments of the distribution as coming from summing over
Green's functions with many external legs which are very susceptible to
fluctuations.  Hence, major reductions in the autocorrelation time for the
magnetization may involve throwing away intermediate measurements of higher
moments which were already effectively independent.  To take the simplest case,
if the different algorithms thereby become effectively comparable and the only
thing that matters is the number of measurements, then the SU(2) runs have
about two or three times the statistics of the $\phi^4$ runs.  If the sytematic
errors were about the same size in both cases, this would mean the SU(2) fit
would have a $\chi^2$ per degree of freedom about two or three times as large,
which is consistent with the observed difference.  These considerations also
tie in with my observations (space considerations forbid extensive reporting of
such details) that bin to bin correlations in the histograms are actually very
slight, even where magnetization auto-correlation times are large (this is why
the individual histograms, as opposed to all of them together, could be fit
with a statistically satisfactory $\chi^2$ even though the error bar estimate
had assumed independence).
I do not presently understand whether these arguments imply that reducing
critical slowing down in the magnetization measurements is never useful in the
context of finite size surfaces, or if they only mean that lattice sizes must
be that much larger before it becomes useful.  Other workers have also observed
an apparent absence of critical slowing down for the probability
distributions\cite{finscal:first_engprobs}.  Certainly, non-local algorithms
that have lower auto-correlation times but need more cpu time should be used
with caution in this context.

The last thing to do is to adjust the scales of one of the plots so that it
fits the other as nearly as possible.  Two operations can be performed.
Firstly, $z$ can be scaled by $z \rightarrow \sigma z$; secondly, $\phi$ can be
changed in a similar manner, $\phi \rightarrow \rho \phi$, but the height of
the surface must be adjusted down by $\ln{\rho}$ to preserve the normalisation
of the probability distributions from which the surface $S_Q$ is obtained.  The
choice is arbitrary, but I decided to scale the $\phi^4$ surface to the SU(2)
one.  These transformations are understood to be acting on the arguments of the
fit function $g(z,\tilde{\phi})$.  Thus if $\sigma$ is 2, then the portion of
the $\phi^4$ surface which did appear on a graph at $z=3$ now appears at
$z=1.5$.

A rigorous statistical analysis of the situation is not possible because of the
uncertain nature of the errors in the data and the uncertainties added by the
inevitable imperfections of the fits.  I adopted a reasonable method which
should be semi-quantitative as to whether the surfaces do match or not. The way
in which I did this was to define a figure of merit, ${\cal R}(\sigma, \rho)$,
for the agreement of the two surfaces, and then minimize this figure.  ${\cal
R}$ was obtained as follows.  First, the $(z,\phi)$ plane was divided up into
rectangles which were small enough that the surface could be considered
approximately linear over them.  Those rectangles which (approximately)
coincided with the regions where data was available to constrain the fit
surface were used; the rest were not.

For each rectangle, the mean square distance of the data in that rectangle from
the surface was separately computed for both models.  This distance was taken
as a measure of the error in the surface over that square.  In so far as the
discrepancies between data and surface were due to systematic errors (either in
the fit or because of slight corrections to finite size scaling between the
histograms), this is a fair measure of the error.  If the process were working
ideally (huge lattices, many histograms with much data, perfect fit function,
small rectangles) then the errors of the data from the fit surface would be
entirely random.  In such a case my error measure for each rectangle would be
an overestimate and should be divided by the square root of the number of data
in the rectangle.  Based on the small lattices involved, inspection of pictures
like Figure \ref{fig:phi4_hist}, and particularly the $\chi^2$ of the fits to
the data, I expected the errors to be predominantly of the sytematic kind, not
random.  The two errors for each rectangle were then combined in quadrature to
give an expected error for the difference between the surfaces.

Finally, the actual mean square difference of the fit surfaces in each
rectangle was computed numerically, and divided by the combined error just
described.  The figure of merit ${\cal R}(\sigma,\rho)$ was then the root mean
square of the individual figures of merit for all the rectangles.  This number
should be less than about one everywhere, and could be considerably less than
one in the ideal case.  It measures the extent to which the surfaces differ by
more than the fluctuations in the data.  The scalings were then obtained by
minimising this function of merit.  They were in good agreement with estimates
obtained by eyeball scaling of the finite size scaling forms of the
susceptibility to each other (about 3\% different).  The actual values were,
for the $z$ scaling $\sigma = 2.631$, and for the $\phi$ scaling $\rho =
1.359$.

Several graphs comparing the SU(2) surface with the rescaled $\phi^4$ surface
are in Figures \ref{fig:view1},\ref{fig:view2}, and \ref{fig:contours}.  I
believe the reader will agree with me that the similarity is very striking,
particularly in the contour plots, Figure \ref{fig:contours}.  As to a more
quantitative understanding of the situation:  a density plot of the merit
figure defined above is to be found in Figure \ref{fig:ratio}.  The numbers on
which it is based are in Table \ref{tab:merit_ratios}.  The root mean square
${\cal R}$ value was $0.50$.  Pretty much all the values for the individual
squares are within the  reasonable range.  I think this makes it clear that the
surfaces for the two models are indeed the same, within the errors.


\section{Conclusion and Outlook}
\label{sec:conclusion}

In this paper I have argued that by taking histograms of the instantaneous
magnetization at different reduced temperatures and lattice sizes and rescaling
them in the appropriate way we can produce smooth finite size scaling surfaces.
 I have further argued that these surfaces should be universal up to two
rescalings.  I have exhibited data for two different models that provide
credible evidence that the histograms can indeed be scaled into surfaces.  I
have further shown that, within reasonable error bars, the two surfaces are the
same.  This conclusion can be looked at in two ways.  If we believe that SU(2)
and $\phi^4$ are in the same universality class, then this work can be seen as
evidence that the finite size surfaces really are universal quantities.  On the
other hand, if we accept the renormalization group arguments for this
universality, the paper provides further support for the view that the
effective line theory for SU(2) lattice gauge theory is in the $\phi^4$/Ising
universality class.  The mutual consistency of the views lends credence to
both.

Several further lines of enquiry could be pursued.  It would be of interest to
know what connection could be made with perturbative expansions.  Some work has
been done to compute the surface of interest
here.\cite{finscal:brezin,finscal:chiral_pert,finscal:helmholtz}  Indeed an
explicit comparison of the probability form at the critical point ($z=0$) in
simulations with an $O(\epsilon)$ result has been done\cite{finscal:helmholtz},
but no comparison exists yet for the form as it varies with temperature.

There should be a similar surface obtained by finite size scaling histograms of
measurements of the internal energy density in a simulation.  These would have
the average internal energy as their expectation value and their variance would
be related to the specific heat.  This may be of interest because of the
numerically small value of the specific heat exponent $\alpha$.  This is very
hard to measure accurately in a simulation.  It being hard to extract from the
data is precisely the condition needed to be able to construct the surface
without knowing $\alpha$ well.  Thus it may be possible to use this to check
universality in the energy variables, where it is difficult to do with an
exponent.

It would be of interest to to know whether the scheme works as well in other
universality classes as it does here.  It would be interesting to see how
different the pictures look from those in Figure \ref{fig:contours}.

If it proves that the scheme does work well elsewhere then it could be quite
useful in situations in which a researcher wishes to be sure of the
universality class of a model but only has Monte-Carlo data to use.  Besides
the exponents, it was previously necessary to compare either amplitude ratios
of one kind or another, or finite size scaling forms of the magnetization,
susceptibility etc.  Firstly, all of these quantities are contained in this
finite size surface (or in other similar ones).  For instance, the
susceptibility amplitude ratio comes from how the `valley' at one end of the
surface narrows in relation to how the one at the other end of the surface
narrows.  Secondly, most of these quantities are not very well suited to
determination from simulation data.  It is hard to determine amplitudes
accurately because the region in temperature between where the correlation
length becomes long enough to be considered in the critical regime and where
finite size effects become pronounced is very small if lattices are only a few
tens of spacings on a side.  It was just this problem which inspired the author
to research the method presented here.  Other authors have had similar
problems. \cite{finscal:landau2D,finscal:landau3D}  As to the shape functions
of the magnetization or susceptibility, they have the drawback of being
uninteresting in the region of the critical point.  The magnetization function
is close to linear, while the susceptibility function is close to quadratic
\cite{finscal:eng90}.  Thus, while cross-model comparison of them is not
totally unimpressive, it is not as convincing as it might be.  Again of course,
these functions are just various moments (taken in the $\phi$ direction) of the
surface described here.  On the other hand, the finite size surface does have
pronounced features in the region of the critical point, and therefore makes
for a test of universality which is both very visual and quite convincing.

Against this it must be said that the technology for the quantitative
comparison of the two surface used here is a little clumsy.  The best hope of
improving this situation is to take some account of the leading corrections to
scaling.  The correction to scaling function and exponent are also universal,
and Bruce has had considerable success in matching the critical point
probability distribution of two apparently disparate models by making use of
this\cite{finscal:bruce_bdr2}.  The usual problem with invoking corrections
to scaling--that they simply provide enough extra parameters that things can be
made to fit no matter what--is somewhat offset by the requirement of using a
single correction function for both models.  Whether this would sufficiently
reduce the systematics that a proper statistical treatment of the random errors
could be given is open to question.

\vspace{0.5cm}

{\bf Acknowledgements:} I am very grateful to Juergen Engels for providing
computer files of the histograms for SU(2) used in this paper.  I wish to thank
Richard Scalettar and Rajiv Singh for helpful discussions and answers to
questions.  The simulations were done over several months of running on a
number of Unix workstations in the Department here and I appreciate the
indulgence of the owners of the various machines.  This research was partially
supported by the Department of Energy.  Finally, Joe Kiskis made important
contributions in the early stages of this project and provided essential advice
and guidance throughout.


\begin{table}[h]
\centering
\begin{tabular}{|r|r|r|r|}   \hline
$N$     &  $r$        &      $\Nrun$ & $z_\phi$  \\ \hline \hline
 32   &  4.1500   &     41347  &   -5.680  \\ \hline
 16   &  4.0000   &     49985  &   -4.769  \\ \hline
 16   &  4.0500   &     99980  &   -3.809  \\ \hline
 16   &  4.1000   &     49975  &   -2.850  \\ \hline
 32   &  4.2000   &     49900  &   -2.797  \\ \hline
 16   &  4.1250   &     99965  &   -2.370  \\ \hline
  8   &  3.9000   &     49990  &   -2.226  \\ \hline
 16   &  4.1500   &     99960  &   -1.890  \\ \hline
  8   &  4.0000   &     49985  &   -1.587  \\ \hline
 16   &  4.1750   &     99960  &   -1.410  \\ \hline
  8   &  4.0500   &     49985  &   -1.268  \\ \hline
 32   &  4.2300   &     57466  &   -1.067  \\ \hline
  8   &  4.1000   &     49985  &   -0.948  \\ \hline
 16   &  4.2000   &     49950  &   -0.931  \\ \hline
  8   &  4.1500   &     49980  &   -0.629  \\ \hline
 16   &  4.2250   &    119642  &   -0.451  \\ \hline
  8   &  4.2000   &     49980  &   -0.310  \\ \hline
  8   &  4.2500   &     49980  &   0.010  \\ \hline
 16   &  4.2500   &     99925  &   0.029  \\ \hline
 32   &  4.2500   &     49700  &   0.086  \\ \hline
  8   &  4.3000   &     49975  &   0.329  \\ \hline
 16   &  4.2750   &     99920  &   0.509  \\ \hline
  8   &  4.3500   &     49975  &   0.648  \\ \hline
 32   &  4.2600   &     49670  &   0.663  \\ \hline
  8   &  4.4000   &     49975  &   0.968  \\ \hline
 16   &  4.3000   &     49900  &   0.988  \\ \hline
 32   &  4.2700   &     49700  &   1.240  \\ \hline
 16   &  4.3250   &     99930  &   1.468  \\ \hline
 32   &  4.2800   &     49800  &   1.816  \\ \hline
 16   &  4.3500   &     99940  &   1.948  \\ \hline
 32   &  4.2900   &     49840  &   2.393  \\ \hline
\end{tabular}

\caption{Summary of $\phi^4$ data set.}
\label{tab:phi4_data}
\end{table}

\begin{table}[h]
\centering
\begin{tabular}{|r|r|r|r|}   \hline
$N$     &  $4/g^2$        &      $\Nrun$ & $z_L$  \\ \hline \hline
  26    &       2.27  &   150000   &  -2.185 \\ \hline
  18    &       2.27  &   100000   &  -1.219 \\ \hline
  26    &       2.29  &   150000   &  -0.652 \\ \hline
  12    &       2.27  &   200000   &  -0.640 \\ \hline
  18    &       2.29  &   300000   &  -0.363 \\ \hline
   8    &       2.27  &   200000   &  -0.336 \\ \hline
  12    &       2.29  &   400000   &  -0.191 \\ \hline
   8    &       2.29  &   200000   &  -0.100 \\ \hline
   8    &       2.30  &   200000   &   0.018 \\ \hline
  12    &       2.30  &   300000   &   0.034 \\ \hline
  18    &       2.30  &   450000   &   0.064 \\ \hline
  26    &       2.30  &   200000   &   0.115 \\ \hline
   8    &       2.31  &   400000   &   0.136 \\ \hline
  12    &       2.31  &   300000   &   0.258 \\ \hline
  18    &       2.31  &   300000   &   0.492 \\ \hline
   8    &       2.35  &   200000   &   0.608 \\ \hline
  26    &       2.31  &   150000   &   0.882 \\ \hline
\end{tabular}

\caption{Summary of SU(2) data set.}
\label{tab:su2_data}
\end{table}

\begin{table}[h]
\centering
\begin{minipage}[t]{3.0in}
\centering
\begin{tabular}{|r|r|}    \hline
a0 & 	4.702    \\ \hline
a1 & 	4.457    \\ \hline
a2 & 	0.029    \\ \hline
a3 & 	0.021    \\ \hline
b0 & 	49.298   \\ \hline
b1 & 	-6.109   \\ \hline
b2 & 	1.182    \\ \hline
c0 & 	426.811  \\ \hline
c1 & 	41.246   \\ \hline
c2 & 	-3.947   \\ \hline
d0 & 	-0.378   \\ \hline
d1 & 	-0.938   \\ \hline
d2 & 	-0.278   \\ \hline
d3 & 	-0.046   \\ \hline
d4 & 	-0.003   \\ \hline
\end{tabular}
\caption{Coefficients of $\phi^4$ surface.}
\label{tab:phi4_coeff}
\end{minipage}
\begin{minipage}[t]{3.0in}
\centering
\begin{tabular}{|r|r|}    \hline
a0 & 	9.249    \\ \hline
a1 & 	20.719   \\ \hline
a2 & 	-3.266   \\ \hline
a3 & 	0.096    \\ \hline
b0 & 	173.632  \\ \hline
b1 & 	-60.028  \\ \hline
b2 & 	-21.462  \\ \hline
c0 & 	2336.416 \\ \hline
c1 & 	1818.542 \\ \hline
c2 & 	-735.435 \\ \hline
d0 & 	-0.122   \\ \hline
d1 & 	-2.404   \\ \hline
d2 & 	-1.559   \\ \hline
d3 & 	-0.547   \\ \hline
d4 & 	-0.072   \\ \hline
\end{tabular}
\caption{Coefficients of SU(2) surface.}
\label{tab:su2_coeff}
\end{minipage}
\end{table}

\begin{table}[h]
\centering
\begin{tabular}{|r|r|r|r|r|r|r|r|r|r|r|r|r|r|r|}   \hline
$\phi$ \verb+\+$z$ & -2.2 & -2.0 & -1.7 & -1.5 & -1.3 & -1.0 & -0.8 & -0.6 &
-0.3 & -0.1 & 0.07 & 0.30 & 0.54 & 0.77 \\ \hline
0.00 & 1.49 & 1.65 & 0.00 & 0.84 & 0.76 & 0.42 & 0.16 & 0.33 & 0.57 & 0.42 &
0.19 & 0.19 & 0.45 & 0.70 \\ \hline
0.09 & 0.60 & 0.46 & 0.00 & 0.42 & 0.58 & 0.25 & 0.11 & 0.36 & 0.59 & 0.36 &
0.16 & 0.18 & 0.44 & 0.61 \\ \hline
0.18 & 0.35 & 0.68 & 0.00 & 0.66 & 0.22 & 0.09 & 0.10 & 0.27 & 0.43 & 0.37 &
0.15 & 0.20 & 0.52 & 0.71 \\ \hline
0.27 & 0.53 & 1.09 & 0.00 & 1.21 & 0.63 & 0.29 & 0.13 & 0.15 & 0.24 & 0.28 &
0.13 & 0.24 & 0.50 & 0.69 \\ \hline
0.36 & 0.44 & 0.82 & 0.00 & 0.85 & 0.55 & 0.24 & 0.09 & 0.09 & 0.09 & 0.13 &
0.06 & 0.18 & 0.43 & 0.75 \\ \hline
0.44 & 0.00 & 0.00 & 0.00 & 0.55 & 0.32 & 0.09 & 0.14 & 0.41 & 0.31 & 0.22 &
0.05 & 0.15 & 0.36 & 0.72 \\ \hline
0.53 & 0.00 & 0.00 & 0.00 & 0.00 & 0.11 & 0.16 & 0.38 & 0.58 & 0.71 & 0.59 &
0.23 & 0.06 & 0.22 & 0.53 \\ \hline
0.62 & 0.00 & 0.00 & 0.00 & 0.00 & 0.00 & 0.00 & 0.41 & 0.75 & 0.80 & 0.82 &
0.42 & 0.13 & 0.14 & 0.34 \\ \hline
0.71 & 0.00 & 0.00 & 0.00 & 0.00 & 0.00 & 0.00 & 0.00 & 0.69 & 0.69 & 0.85 &
0.33 & 0.17 & 0.28 & 0.26 \\ \hline
0.80 & 0.00 & 0.00 & 0.00 & 0.00 & 0.00 & 0.00 & 0.00 & 0.00 & 0.69 & 0.57 &
0.31 & 0.13 & 0.28 & 0.42 \\ \hline
0.89 & 0.00 & 0.00 & 0.00 & 0.00 & 0.00 & 0.00 & 0.00 & 0.00 & 0.00 & 0.00 &
0.24 & 0.07 & 0.24 & 0.43 \\ \hline
0.98 & 0.00 & 0.00 & 0.00 & 0.00 & 0.00 & 0.00 & 0.00 & 0.00 & 0.00 & 0.00 &
0.00 & 0.00 & 0.00 & 0.36 \\ \hline
1.07 & 0.00 & 0.00 & 0.00 & 0.00 & 0.00 & 0.00 & 0.00 & 0.00 & 0.00 & 0.00 &
0.00 & 0.00 & 0.00 & 0.00 \\ \hline
1.16 & 0.00 & 0.00 & 0.00 & 0.00 & 0.00 & 0.00 & 0.00 & 0.00 & 0.00 & 0.00 &
0.00 & 0.00 & 0.00 & 0.00 \\ \hline
\end{tabular}
\caption{Merit ratios used in Fig \protect\ref{fig:ratio}.  $\phi/z$ values
given here are for the lower left corner of the square shown in the figure
(over which the merit ratio figure is an average).}
\label{tab:merit_ratios}
\end{table}

\begin{figure}[h]
\caption{$\phi^4$ histograms for the rescaled field $\phitil$, as described in
the text, plotted versus $z_\phi$, the finite size scaled reduced temperature.
The tendency of the data curves to all lie on a surface is clear.}
\label{fig:phi4_hist}
\end{figure}

\begin{figure}[h]
\caption{$\phi^4$ scaling surface, $S_Q(z_\phi,\phitil)$, with associated data.
 The data is minus the logarithm of the histograms of Figure
\protect\ref{fig:phi4_hist}.}
\label{fig:phi4_data}
\end{figure}

\begin{figure}[h]
\caption{SU(2) scaling surface, $S_Q(z_L,\tilde{L})$, with associated data.
Agreement is generally fairly good except for one tail where the simulation ran
up to unusually high field.}
\label{fig:su2_data}
\end{figure}

\begin{figure}[h]
\caption{SU(2) scaling surface and rescaled $\phi^4$ scaling surface.  The
$\phi^4$ surface has been rescaled to match the SU(2) surface, so both are
given as functions of $z_L$ and $L$.  See the text.}
\label{fig:view1}
\end{figure}

\begin{figure}[h]
\caption{SU(2) scaling surface and $\phi^4$ scaling surface, a second view.
The $\phi^4$ surface has been rescaled to match the SU(2) surface, so both are
given as functions of $z_L$ and $L$.  See the text.}
\label{fig:view2}
\end{figure}

\begin{figure}[h]
\caption{Contour plots of the SU(2) scaling surface and $\phi^4$ scaling
surface.  The $\phi^4$ surface has been rescaled to match the SU(2) surface, so
both are given as functions of $z_L$ and $L$.  The contours run from 7.0 down
in steps of 0.5.}
\label{fig:contours}
\end{figure}

\begin{figure}[h]
\caption{Figure of merit for comparison of the two surfaces, as a function of
$z_L$ and $L$ (described in text - see also Table
\protect\ref{tab:merit_ratios}).  Black squares represent merit figures of one
or worse.  Values very close to zero are represented by light colours.  White
squares mean that no data was available in that square to normalize the
discrepancy between the two finite size surfaces.}
\label{fig:ratio}
\end{figure}

\end{document}